\documentclass[11pt]{article}

\usepackage[margin=1in]{geometry}
\usepackage{amsmath,amssymb,amsthm}
\usepackage{graphicx}
\usepackage{hyperref}
\hypersetup{
    hypertexnames=false,
    hidelinks,
    pdftitle={RANDAO Manipulation in the Presence of MEV},
    pdfauthor={Kaya Alpturer, Nicholas Hope, S. Matthew Weinberg}
}
\usepackage[nameinlink,capitalise]{cleveref}

\usepackage{enumerate}
\usepackage{float}
\usepackage{tikz}
\usetikzlibrary{
    arrows,
    automata,
    positioning,
    shapes.geometric,
    arrows.meta,
    positioning
}
\usepackage[breakable,most]{tcolorbox}
\newtcolorbox{box1}[2][]{
    breakable,
    colback=white,
    colframe=black,
    boxrule=0.5pt,
    colbacktitle=black,
    coltitle=white,
    enhanced,
    left=2mm,
    right=2mm,
    top=2mm,
    bottom=2mm,
    title={#2},
    #1
}
\usepackage{pgfplots}
\pgfplotsset{compat=1.18}
\usetikzlibrary{calc}

\definecolor{modelUniform}{RGB}{31,119,180}
\definecolor{modelPerSlot}{RGB}{255,127,14}
\definecolor{modelSmooth}{RGB}{214,39,40}
\definecolor{modelThreshold}{RGB}{44,160,44}
\pgfplotsset{
    paperplot/.style={
        width=0.8\linewidth, height=5cm,
        legend cell align=left,
        legend style={font=\footnotesize, draw=gray!40, fill=white,
                      fill opacity=0.92, text opacity=1, rounded corners=2pt,
                      inner sep=3pt},
        label style={font=\small},
        tick label style={font=\small},
        title style={font=\small},
        grid=major,
        grid style={dashed, gray!25},
        axis line style={gray!50, line width=0.5pt},
        tick style={gray!50},
        every axis plot/.append style={thick, line cap=round, line join=round,
                                       mark options={fill opacity=1, scale=0.85}},
    },
    percent x/.style={xticklabel={\pgfmathparse{\tick*100}\pgfmathprintnumber[fixed, precision=0]{\pgfmathresult}\%}},
    percent y/.style={yticklabel={\pgfmathprintnumber[fixed, precision=0]{\tick}\%}},
}

\newcommand{\N}{\mathbb{N}} \newcommand{\R}{\mathbb{R}}
\newcommand{\E}{\mathbb{E}}

 \newcommand{\dist}{\mathcal{E}}
 \newcommand{\tail}{\mathsf{tail}}

\newcommand{\nontp}{\mathsf{non\text{-}tail\text{-}proposals}}
\newcommand{\propose}{\mathsf{propose}} \newcommand{\nnext}{\mathsf{next}}
\newcommand{\OPT}{\mathsf{OPT}}

\renewcommand{\eth}{\mathtt{ETH}}

\usepackage{xcolor}
\usepackage{booktabs}
\usepackage[bibliography=common,appendix=append]{apxproof} %
\newtheoremrep{theorem}{Theorem}
\newtheoremrep{lemma}[theorem]{Lemma}
\theoremstyle{definition}
\newtheorem{definition}[theorem]{Definition}
\newtheorem{example}[theorem]{Example}

\title{RANDAO Manipulation in the Presence of MEV}

\newcommand{\paperacknowledgements}{This work was supported by an Ethereum Foundation grant.
During Prof.\ Weinberg's development of this paper, he participated as an expert witness on
behalf of the State of Texas in ongoing litigation against Google (the ``Google Litigation'').}

\author{
    Kaya Alpturer
    \thanks{Princeton University, \texttt{kalpturer@princeton.edu}.}
    \and Nicholas Hope
    \thanks{Talarion, \texttt{nick@talarion.tech}.}
    \and S.\ Matthew Weinberg
    \thanks{Princeton University, \texttt{smweinberg@princeton.edu}.}
}
\date{}

\begin{document}
\maketitle
\begin{abstract} %
Ethereum's randomness beacon (RANDAO) is well-known to be manipulable,
and prior work \cite{AW24} computes the precise fraction of blocks a
strategic proposer can propose.
The fraction of blocks proposed, however, is only a proxy for participants' rewards.
We propose a generalized reward model capturing many canonical forms of rewards:
consensus reward rollover, multi-block MEV, CEX-DEX arbitrage, oracle manipulation,
and others.
We provide a methodology that computes an $\varepsilon$-optimal strategy
for any reward scheme in our model (and in particular,
any combination of the above rewards).
Finally, we apply our methodology to several canonical examples,
and establish the sensitivity of RANDAO manipulation to the underlying rewards.
We find that if rewards partially roll over, or scale super-linearly with
consecutive blocks, the incentive to manipulate RANDAO is amplified.
Lastly, we investigate tail-slot slashing, which can be modeled as a
reward function, and show that honest equilibria can be recovered.
 \end{abstract}

\section{Introduction}
A major concern for blockchain protocols is the potential for economically
rational participants to deviate from the protocol specification to
maximize their rewards.
In many protocols, randomness is used to ensure the
fair distribution
of block proposing rights.
Thus, if the randomness in a protocol is not robust,
deviations that bias it will be incentivized.
In Bitcoin, protocol deviations such as selfish mining have
relied mainly on forking and hiding blocks, as biasing the randomness requires
breaking the robust SHA256 hash function.
In Proof-of-Stake protocols,
however, the randomness is often
provided by a distributed randomness beacon built into the protocol itself, creating a novel attack vector.
In particular, Ethereum's implementation of randomness,
called RANDAO, has been well
known to be vulnerable to block withholding attacks \cite{AW24}.
Given RANDAO's role in
determining the winners of the block proposal lottery,
there is great financial
incentive to manipulate it.

With Ethereum's stake pool at roughly \$80 billion, the concern for RANDAO
manipulation has resulted in several lines of work, analyzing the potential
impacts of this type of deviation \cite{AW24,fork25,slots25,ethresearchblog}.
These works establish that an adversary with an $\alpha$ fraction of the total
stake is able to propose strictly more than an $\alpha$ fraction of the total
blocks, and therefore earn greater than an $\alpha$ fraction of the total
consensus rewards.

However, rewards in Ethereum have increasingly been dominated by
execution layer rewards that stem from decentralized finance
activity as opposed to constant rewards per block proposed.
We have seen, for example, that sophisticated actors are able to
extract rewards from arbitrage, front-running, and transaction fees
\cite{daian2020flash,OSTM24}.
Therefore, there is a need to model rewards more generally to include
higher-order reward properties beyond a single block.

The importance of varying rewards is
well-established in the proof-of-work setting.
For Bitcoin, \cite{CKWN16} shows that
variance in transaction fees fundamentally
alters the equilibrium structure of selfish mining, with subsequent work
generalizing the analyses to richer reward models~\cite{BNW25,ZAET22}.
For RANDAO manipulation, however, the analogous question of how MEV and
similar non-uniform reward structures affect the incentive to
manipulate it has remained open. While prior work laid the foundation for
the MDP-based analysis and formalization of the larger problem,
these analyses mainly measure the fraction
of blocks proposed.

\textbf{Our primary contribution is a sampling-based methodology to compute
	$\varepsilon$-optimal policies for RANDAO manipulation for significantly
	more general reward models.}
Our approach relies on a novel truncation of the RANDAO game that can provably
provide bounds on the optimal solution.
Prior work \cite{AW24} relies critically on uniform rewards
to collapse the state space, and our truncation and sampling-based
approach requires a new argument for bounding the optimal manipulation.

As a warmup exercise, we show that our methodology reproduces the results
of \cite{AW24} in the case of uniform rewards (single unit of reward per proposed block).
Our approximate solver is able to more efficiently compute approximate results, and
larger runs with sufficient samples are capable of achieving progressively more accurate
results.
We then investigate several reward models relevant to transaction fees,
arbitrage, oracle manipulation, and consensus rewards.
In the process, we characterize specific types of MEV and their
interactions with RANDAO manipulation, including
\emph{per-block rewards (with partial rollover)},
and two models of \emph{multi-block MEV}.

\subsection{Overview of Methodology}

Below, we overview several instances of execution layer rewards, and how they fit into our methodology. Recall that RANDAO proceeds as follows. Every time a proposer proposes a block during their slot, RANDAO changes, becoming a new value that is: (a) deterministic, to the proposer, (b) appears uniformly random, to anyone without the proposer's private key. At the end of every 32-block epoch, the next 32 proposers are selected according to the current RANDAO value.\footnote{In the actual Ethereum implementation, the RANDAO from epoch $i$ determines the proposers for epoch $i+2$. We ignore interleaving and discuss its effect in the model assumptions.} Therefore, when proposers are selected for the last $k$ blocks of an epoch, they can selectively propose/skip slots to select among $2^k$ independent random options for the next epoch's 32 proposers. The cost of doing so is whatever rewards are sacrificed from skipped blocks. The benefit of doing so is the potential for more rewards (and perhaps more opportunities for future manipulation) in the next epoch.
We call the largest such $k$ controlled by an adversary the \emph{tail}.

When discussing sample rewards below, we use \textsc{Tail}
to refer to the pattern of proposed/skipped slots by the adversary in the
tail of an epoch and \textsc{Next} for the pattern of proposed slots by the adversary in the non-tail.
Assuming that other proposers honestly follow the intended protocol (and propose in every slot), a strategic proposer who does not control the last slot of an epoch optimizes their expected rewards by proposing in every non-tail slot. This is simply because: (a) such a proposer has no control over the next epoch's RANDAO, (b) a proposer seeking to optimize rewards in this epoch alone can always replace a skipped block with an empty block without affecting the execution rewards and still collect block rewards.

\begin{example}[Reward Rollover] \label{ex:perslot}
	Many execution layer rewards, such as those from on-chain arbitrage and sandwich attacks, roll over
	if not claimed in a block. For instance, an MEV opportunity in the mempool simply becomes available to the next proposer
	after a missed block.\footnote{See \cref{fig:bid_histogram} for the distribution of MEV bids on the MEV-Boost auction showing the distribution of these rewards.}
	This also occurs with consensus layer rewards. The block reward increases after a miss due to the increase in
	the number of attestations that can be included (see \cref{fig:cl_rewards}).
	Intuitively, this rollover helps RANDAO manipulation by reducing the cost of
	missing a block if the adversary knows that they will get to
	propose after a miss. Because this rollover can be represented as a function on $\textsc{Tail}$ and $\textsc{Next}$, we investigate its
	effect in our per-slot reward model.
\end{example}

\begin{example}[Arbitrage Value] \label{ex:smooth}
	For an adversary controlling $k$ consecutive blocks, a single arbitrage at the end is more profitable than multiple arbitrage transactions
	due to trading fees \cite{milionis2024automated}.
	Fees introduce a profitability threshold, giving the arbitrage an option-like
	payoff.\footnote{For instance, consider a scenario in which DEX and CEX prices begin perfectly equal, with a de minimis fee parameter on the CEX but a 30 basis point fee on the DEX. In such a scenario, a price discrepancy of more than 30 basis points must arise via price movement on the CEX before a validator may execute an arbitrage transaction profitably. Reserving sole arbitrage rights via multi-slot control thus exhibits the same nonlinear payoff function as a put or call option: the opportunity is worthless until the option becomes in the money (that is, the price discrepancy exceeds cumulative fee parameters of the target CEX-DEX pair).}
	Controlling a longer sequence of $k$ blocks increases the probability that the price will drift beyond this threshold.
	Thus, a validator intending to perform CEX-DEX arbitrage values
	a run of $k$ consecutive blocks more than $k$ independent blocks,
	motivating a superlinear model for MEV extraction.
	In particular, observe again that the reward earned by the proposer can be written as a function on \textsc{Tail} and \textsc{Next}.
\end{example}

\begin{example}[Oracle Manipulation] \label{ex:threshold}
	Manipulation of time-weighted average price (TWAP) oracles in order to force
	illegitimate liquidations is a well-understood form of multi-block MEV \cite{mackinga2022twap}.
	While oracle manipulation of this form is already a concern without an adversary controlling multiple blocks in a row\footnote{Prior work \cite{mackinga2022twap} primarily considers block proposers that include the attacker's transactions, but also transactions from arbitrageurs that de-manipulate asset prices. Such a model remains appropriate in a setting where the sequence of proposers cannot be guaranteed in advance, such as Proof-of-Work in contrast to Ethereum.},
	RANDAO manipulation exacerbates the threat by allowing an adversary
	complete control over transaction inclusion, reducing the attack cost. For a given adversary, TWAP attacks represent a model in which a large, fixed value (the liquidation profit) is available if the validator can secure enough sequential blocks to initiate the liquidation.
	As such rewards can be written as a function on \textsc{Tail} and \textsc{Next}, we explore this further in the threshold multi-block MEV model.
\end{example}

\subparagraph*{Our Methodology.} All of the above examples can be captured by specifying two functions $f(\cdot)$ and $g(\cdot)$. $f(\cdot)$ takes as input a sequence of proposals by the adversary in the non-tail part of the epoch (which we refer to as $\textsc{Next}$),
and outputs ``under the specified reward model, what is the expected reward for the adversary if the adversary proposes all of their blocks and honest validators propose the other blocks?'' $g(\cdot)$ takes as input a fully specified $\textsc{Tail}$ sequence and outputs ``under the specified reward model, what is the expected reward for the adversary if the adversary proposes the specified blocks and misses the others in \textsc{Tail}?''
Note the distinction between the two: in $\textsc{Next}$ the non-proposed blocks are proposed by an honest party and in
$\textsc{Tail}$ non-proposed blocks are missed.
As we later explore, this may make a difference if honestly proposed blocks affect
rewards of other blocks.
For \emph{any} reward model that can be fully specified by some $f(\cdot)$ and $g(\cdot)$ (including all three examples above, and any combination of them), our methodology computes an $\varepsilon$-optimal strategy for a strategic proposer.
In the formal model of \cref{sec:model}, $f$ corresponds to $\nnext(\cdot)$ and
$g$ corresponds to $\propose(\cdot)$.

\subparagraph*{Highlights.}
In addition to developing our methodology,
we also apply it to explore canonical settings and draw conclusions about the role of execution rewards
in RANDAO incentives. First, we
reproduce the results of \cite{AW24} in the case of uniform rewards (constant per block),
and show that our approximate solver can quickly reproduce these
results at coarse $\varepsilon$.
We then demonstrate the methodology on three execution-layer
reward models: per-slot rollover, a smooth multi-block model,
and a threshold multi-block model. Consensus-reward rollover is directly
visible in the data (\cref{fig:cl_rewards}), but MEV rollover is harder
to verify empirically: slot misses are rare on Ethereum, so the
available observational data on what happens to MEV opportunities
across a missed slot is scarce. Systematic
multi-block MEV is also weak in current Ethereum~\cite{stichler24},
so the smooth and threshold models are stylized. The modelled MEV
regimes may matter in the future under block-building redesigns.
In the regimes we model, partial reward rollover and multi-block MEV
both amplify the incentive to manipulate RANDAO.

Lastly, we focus on two practically relevant investigations using the same
framework. We show how increasing the epoch length decreases the
manipulation gains across reward models\footnote{\cite{AW24} observed this
	for uniform rewards. Our results extend it to non-uniform reward structures,
	strengthening the case for longer epochs as a mitigation.}.
We then encode, as a reward function, a tail-slashing curve that penalizes
misses toward the tail more heavily and amplifies penalties for groups of
consecutive misses, and show that honest behavior can be made
$\varepsilon$-optimal under appropriate slashing magnitudes.

\subsection{Related Work} \label{sec:related}
\subparagraph*{RANDAO Manipulation.}
Since its inception, RANDAO has been known to be manipulable to a certain degree
by withholding tail slots \cite{eth2book,ethresearchblog}. Early probabilistic modeling based
work \cite{AR19} shows that this manipulation is small but non-negligible.
\cite{AW24} develops a Markov Decision Process (MDP) formulation of RANDAO
manipulation to compute optimal strategies. Later work extends this analysis to
consider forking \cite{fork25,slots25}. Aside from RANDAO, prior work also considers manipulation of other proof-of-stake protocols~\cite{CLWZ24,FerreiraHWY22,FerreiraGHHWY24}. However, all prior work considers strategic proposers who seek only to optimize their fraction of proposed slots, and do not consider variable execution rewards.

\subparagraph*{Selfish Mining.}
The analogous manipulation in Bitcoin is selfish mining, first analyzed by
\cite{EyalS14}. Subsequent work improves the analysis and considers optimal
strategies \cite{ZAET22,SapirshteinSZ16,ZET20} via an MDP methodology. Several works consider the effect of transaction fees on selfish mining.
\cite{CKWN16} shows that transaction fees radically change the analysis, and selfish mining variants can be profitable for arbitrarily small miners~\cite{ZAET22}. In a similar spirit to our work, \cite{BNW25}
develops a generalized model of selfish mining to accommodate execution layer rewards.
In comparison to these works, ours is the first to consider execution
rewards such as MEV extracted through multiple consecutive blocks.
However, our model does not subsume these -- a strategic proposer in~\cite{BNW25,CKWN16} can adapt their strategy on the realized draw of transaction fees, whereas a proposer in our model does not.
Moreover, our methodology finds the optimal strategy, rather than analyzing specific strategies that outperform honesty.

\subparagraph*{Computing Optimal Manipulations.}~\cite{SapirshteinSZ16} are the first to use MDP methodology to compute optimal manipulations in a blockchain protocol. MDPs are also used in~\cite{AW24,ZAET22,ZET20}, and similar convergence-based techniques are used in~\cite{FerreiraGHHWY24}. In comparison to these works, ours is the first to find optimal strategies for execution rewards beyond the uniform block rewards. In comparison to~\cite{AW24} specifically, our methodology can estimate $\varepsilon$-optimal strategies more efficiently and accommodate significantly more complex reward structures via sampling.

\subparagraph*{Maximal Extractable Value.}
Maximal extractable value (MEV) has been a large and growing part of reward
modeling \cite{babel2023clockwork,daian2020flash}. Recently, there has been interest in understanding the
game-theoretic implications of MEV on the incentive-compatibility of blockchain
protocols \cite{bahrani2024transaction,mevlimitscaling,schwarz2023time}. MEV's
interaction with various kinds of protocol deviations is an active area of
research.

\subparagraph*{Validator Rewards and Slashing.}
Modifying validator rewards to disincentivize strategic deviations
is a recurring theme in blockchain protocol design. Ethereum currently
penalizes missed attestations uniformly via the inactivity
leak~\cite{eth2book}, and there have been community discussions of
introducing missed-slot penalties as a more targeted
intervention~\cite{missedslot_tempcheck}. A parallel line of work proposes
time-sensitive or retroactive block rewards to disincentivize strategic
delays in block proposal (timing games)
\cite{ABS25,schwarz_retroactive,schwarz2023time}. We investigate a
similar mitigation for RANDAO manipulation: we model tail slashing
penalties as a reward function within our framework, and apply our
methodology to determine slot miss penalties at the tail under which honest
behavior is $\varepsilon$-optimal.

\subsection{Main Contributions}
\begin{itemize}
	\item We analyze RANDAO manipulation under a wide array of stylized
	      reward models (\cref{sec:results}). This improves prior work by: (a)
	      being the first to consider varying rewards such as execution layer
	      rewards, and (b) allowing quick exploratory runs at coarse
	      $\varepsilon$.
	\item We introduce a novel truncation of the RANDAO game that provides
	      provable upper and lower bounds (\cref{sec:mdp}), rendering many
	      instances of the game with complicated reward functions
	      computationally tractable.
	\item We prove robustness guarantees for our methodology, and provide a
	      codebase\footnote{\url{https://github.com/kalpturer/randao-mev}.}
	      for easy replication of our results and running similar analyses
	      for any reward function of interest.
	\item We apply our methodology to canonical settings of interest
	      (\cref{sec:improvement}, \cref{fig:improvement}), and conclude that
	      the effect of RANDAO manipulation is highly dependent on the
	      underlying reward model.
	\item We apply our framework to evaluate practical mitigations:
	      epoch-length variation (\cref{fig:epoch_sweep}) and tail slashing
	      penalties (\cref{sec:mitigation}). Modeling slashing as a reward
	      function within our methodology, we compute, for each MEV reward
	      model, minimal slashing curves that recover honest behavior at
	      the equilibrium.
\end{itemize}

\subsection{Roadmap}
In \cref{sec:model}, we model RANDAO manipulation with general
rewards. In \cref{sec:mdp}, we present our methodology of computing
$\varepsilon$-optimal policies. We then proceed to analyze complex
reward models in \cref{sec:results}.
In \cref{sec:improvement}, we compare different reward models and show how
much better a strategic proposer can do compared to honest proposers,
including a sensitivity analysis on the epoch length.
In \cref{sec:mitigation}, we analyze slashing-based mitigation
approaches under which honest behavior is $\varepsilon$-optimal.
Finally, \cref{sec:conclusion} concludes with a summary and future work.
We defer proofs to the appendix.
\section{Modeling RANDAO} \label{sec:model}
We now model RANDAO manipulation in its full generality, where the reward
function is dynamic.

\subsection{Background}
\subparagraph*{Ethereum Consensus.}
Ethereum uses a Proof-of-Stake consensus mechanism which divides time
into epochs with 32 slots, each of which lasts for 12 seconds. The
participants in the protocol are called \emph{validators} and stake $\ge 32~\eth$
each. For each epoch, 32 validators are sampled uniformly at random
to propose the blocks in the epoch. The sampling procedure requires a reliable
source of randomness, which is provided by the RANDAO mechanism \cite{eth2book}.

\subparagraph*{RANDAO Mixing.}
Each block contains a 256-bit random seed called the RANDAO. The validator
deterministically computes the block RANDAO by signing the current epoch number
and then hashing the resulting signature. This hash is then mixed into the
RANDAO value via an XOR operation. Note that this mixing is verifiable since the
public key of the validator is known and the validator also includes the raw
signature in the block. Thus, there is a unique value which a validator may use
when updating the RANDAO. The future value of the RANDAO nonetheless cannot be
computed by any third party ahead of time, since it relies upon both (a) the
validator's private key, and (b) the RANDAO value as of the preceding block. The
only exception to this rule occurs if and when a sequence of validators collude
to reveal RANDAO values in advance. We model such a group of validators as the
adversarial proposer.

\subparagraph*{Proposer Selection.}
The RANDAO value at the end of epoch $E$ is used to determine the proposers for
the next epoch by deterministically sampling 32 validators. The sampling protocol
is complex, but can be accurately modeled by 32
independently and uniformly random selections from the pool of available
validators.

\subparagraph*{Rewards.}
Ethereum employs a reward mechanism that incentivizes validators to propose,
vote on blocks, and participate in the consensus protocol. Consensus-layer
rewards are predictable and constitute a low-variance incentive for block
proposal.\footnote{The rewards are not fixed, but depend on the total number of
	validators in the network. However, this number is relatively stable.}
Execution-layer rewards, by contrast, are highly stochastic.
The desire to maximize execution-layer rewards has driven
the rise of a specialized and highly sophisticated ecosystem of Maximal
Extractable Value\footnote{Maximal Extractable Value is defined as the
	reward available to a block proposer via their ability to selectively include,
	exclude, insert, and reorder transactions.} (MEV) searchers, block
builders, and relays \cite{OSTM24}.

\subparagraph*{Adversary.}
We assume a single adversary that controls a fraction $\alpha$ of the total
network stake.
All other validators are assumed to be under the control of honest proposers and
will therefore propose in each available slot. The adversary
makes a strategic choice to propose or miss in each slot that they are assigned.
We will hereafter refer to the consecutive proposal slots that the adversary is
assigned at the end of an epoch as the \emph{tail} of the epoch\footnote{For
	example, denoting adversarial slots by `A' and honest slots as `H', ($\dots$H)
	has tail 0, ($\dots$HA) has tail 1, ($\dots$HAA) has tail 2, and so on.}.

\subparagraph*{RANDAO Manipulation.}
Suppose the adversary controls a tail of length $t$ in epoch $E$. Each subset of
slots in which the adversary could propose will produce a unique
RANDAO value at the end of the epoch.\footnote{Note that this is only the case
	because each slot in the tail is associated with a \textit{unique} validator,
	each of which is under the control of the adversary. Thus the associated
	signatures are unique, meaning that the adversary has $t$ distinct $256$ bit
	strings which they may mix into the RANDAO.} Because the adversary controls the
validators assigned to the tail, and hence knows their private keys, the RANDAO
value associated with each slot can be deterministically precomputed. Thus the
adversary may sample $2^t$ distinct future values of the RANDAO corresponding to
each subset of tail slots in which they could propose.

\subparagraph*{Modeling Assumptions.}
We make some assumptions while modeling the RANDAO mechanism:
\begin{enumerate}[(a)]
	\item Proposer sampling is uniformly random and the adversary gets selected to
	      propose in a slot with $\alpha$ probability. Hence, each epoch assignment consists of
	      32 i.i.d. samples.
	\item We ignore interleaving of epochs and assume that the RANDAO value at
	      the end of an epoch determines the proposers for the next epoch.\footnote{There
		      is justification for this assumption in prior work \cite{AW24}, and it
		      essentially does not affect the analysis for the adversarial stake
		      range of interest ($\alpha < 1/2$) since each epoch contains at least
		      one honest proposer with high probability.}
	\item Honest proposers always propose and the adversary always chooses to
	      propose when they are in non-tail slots as it is in their best interest to do so.
	\item We assume that the reward ``resets'' after an
	      honest proposer, i.e., the reward distribution before and
	      after an honest proposer is independent.
\end{enumerate}

\subsection{RANDAO Game}
We now define the game induced by manipulating the RANDAO with an abstract
reward model. Since general rewards may depend on the state in arbitrary ways,
we begin by modeling the state space in its most general form as well. As an
illustrative example, consider a scenario in which the adversary controls a tail
of length $3$. There are $2^3$ different sequences of proposals that the
adversary can make and one option is to do $\langle \mathsf{propose},
	\mathsf{miss}, \mathsf{propose}\rangle$ which suppose results in a future epoch assignment with tail
30 of the form $\langle \mathsf{adversary}, \mathsf{honest}, \mathsf{adversary}, \dots \rangle$. In
a reward model in which we only consider the number of blocks, this action will
have a reward of 3 blocks plus getting to keep playing the game from a tail of
30. In the general case, however, there will be a known reward distribution that
depends on the proposal sequence\footnote{One way to think about this is to
	consider the distribution of rewards that a particular adversary is able to
	obtain. For example, if we consider an adversary that is really good at
	extracting multi-block MEV, the reward distribution will depend on whether
	proposals are consecutive. Alternatively if there is not much correlation
	between block rewards, we can use a per-slot independent model. We consider
	these in more detail in \cref{sec:results}.}, and the adversary will
first evaluate the expected reward of $\langle \mathsf{propose}, \mathsf{miss},
	\mathsf{propose}\rangle$ followed by the expected reward of $\langle
	\mathsf{propose}, \mathsf{honest\; proposal} \rangle$ and the subsequent value of
playing the game from a tail of length 30.

\subparagraph*{Proposal strings.}
At the core of our formulation is the definition of two types of proposal strings.
\begin{itemize}
	\item \emph{tail proposal string $\textsc{(Tail)}$:} $p_t \in \{0,1\}^{k}$ represents the
	      event in which out of $k$ consecutive slots all assigned to the
	      adversary to propose, the adversary proposed in exactly the indices
	      with 1.
	\item \emph{non-tail proposal string $\textsc{(Next)}$:} $p_n \in \{0,1\}^k$ represents the
	      event in which out of $k$ consecutive slots, the adversary proposed
	      in exactly the indices with 1, while honest validators proposed in
	      the remaining indices with 0 (recall that we assume all honest
	      parties always propose).
\end{itemize}
Intuitively, when deciding what option to choose, the adversary will consider
the expected reward of these two events and the resulting new tail of the
action.

\subparagraph*{Epoch assignments.}
Let $\dist$ be the distribution of epoch assignments of the form $\{0,1\}^{32}$
where each bit is identically and independently drawn with probability $\alpha$.
We will refer to the tail length of an epoch assignment $E \sim \dist$ as
$\tail(E)$ and the non-tail proposal string as $\nontp(E)$.

\subparagraph*{Reward model.}
We define the rewards as an arbitrary function of the proposal strings
representing the expected reward of the reward distribution. A reward model
$\mathcal{R}$ is a pair $\langle \mathsf{propose}, \mathsf{next} \rangle$ where
\begin{itemize}
	\item $\mathsf{propose} : \{0,1\}^* \to \R$ is a function that takes in a
	      tail proposal string and outputs the expected reward of the adversary in taking
	      this action.
	\item $\mathsf{next} : \{0,1\}^* \to \R$ is a function that takes in a
	      non-tail proposal string and outputs the expected reward of the adversary in
	      taking this action.
\end{itemize}

\begin{box1}[]{}
	\noindent\textbf{RANDAO game.}
	Now, combining all the observations, we define the RANDAO manipulation game
	parameterized by adversarial stake $\alpha$ and reward model $\mathcal{R}$. The
	variables $\textsc{Reward}$ and $\textsc{Rounds}$ count the accumulated reward
	by the adversary and the number of epochs that have passed. $\textsc{Tail}$ is
	used to keep track of the current tail controlled by the adversary.
	\begin{enumerate}[(a)]
		\item \textit{Initialization.}
		      \begin{enumerate}[1.]
			      \item Let $\textsc{Reward} := 0$ and $\textsc{Rounds} := 0$.
			      \item Let $\textsc{Tail} := 0$.
		      \end{enumerate}
		\item \textit{Sampling.}
		      \begin{enumerate}[1.]
			      \setcounter{enumii}{2}
			      \item Let $E_P \sim \dist$ for all $P \subseteq
				            \{0,1\}^{|\textsc{Tail}|}$.
			      \item Let $N_P$ be $\nontp(E_P)$ and similarly let $T_P$ be
			            $\tail(E_P)$.
		      \end{enumerate}
		\item \textit{Strategic choice.}
		      \begin{enumerate}[1.]
			      \setcounter{enumii}{4}
			      \item The adversary chooses a subset $P^*$ to propose in based
			            on the reward model $\mathcal{R}$ and the set of sampled
			            epochs.
		      \end{enumerate}
		\item \textit{Updates.}
		      \begin{enumerate}[1.]
			      \setcounter{enumii}{5}
			      \item Let $\textsc{Reward} := \textsc{Reward} +
				            \propose_\mathcal{R}(P^*) + \nnext_\mathcal{R}(N_{P^*})$.
			      \item Let $\textsc{Tail} = T_{P^*}$.
			      \item Let $\textsc{Rounds} := \textsc{Rounds} + 1$.
			      \item Repeat from step 3.
		      \end{enumerate}
	\end{enumerate}
\end{box1}

\subparagraph*{Objective.}
The optimization objective is $\lim\inf\{ \textsc{Reward} / \textsc{Rounds} \}$,
the average reward in the infinite game.
\section{Solving the RANDAO Game} \label{sec:mdp}

In this section, we define the MDP formulation of the RANDAO game, using the
state space reduction observation from \cite{AW24}. The MDP formulation reduces
the state space to just the tail lengths $\{0,1,\dots,32\}$, pushing the
complexity of the game into the action space. The MDP solver in \cite{AW24},
however, does not immediately generalize to our case.

Diverging from \cite{AW24}, to give probable bounds on our sampling-based approach,
we observe that the rare states can be truncated without
significantly affecting the optimal reward. Concretely, we will define two
truncations of the MDP where we ignore states above a certain threshold: (a) an
\emph{upper truncation} that rewards large tails state transitions maximally,
and (b) a \emph{lower truncation} that rewards large-tail state transitions 0.
This will allow us to bound the optimal reward of the full MDP between the
optimal rewards of the two truncated MDPs.

\subparagraph*{Comparison with \cite{AW24}.}
Our MDP uses the same state space as \cite{AW24}, the tail length
$t \in \{0,\dots,32\}$. The main difference is in computing the transition
probabilities. Computing each $P(t,t')$ requires summing, over the
reward realizations the adversary could face from $t$, the probability
that the adversary's optimal response leads to $t'$. Under uniform
per-block rewards the sum has manageably few terms because the optimal
response depends only on the \emph{number of proposals}. Once the reward
depends on the \emph{pattern}, as in our examples, the sum has many
more terms and quickly becomes intractable to evaluate.
For example, the smooth multi-block model (Example~\ref{ex:smooth})
pays some function of $k$ for each block of $k$ consecutive proposals, so the
sum has to enumerate over integer partitions of every proposal count
$k \leq t$.
Our methodology sidesteps
this combinatorial issue by estimating the transitions via
sampling, with an $\varepsilon$-approximation guarantee from a
truncation argument and a concentration bound. As a practical
benefit, new reward models can be plugged into the solver without
re-deriving any closed forms.

\subsection{Methodology}
Observe that we can immediately write down the MDP for the RANDAO game with
reward model $\mathcal{R}$. Let $\mathcal{S} = \{0,1,\dots,32\}$ be the state
space, $\mathcal{A}$ be the set of valuation functions
$v : S \to \R$.
Denote by $\pi : \mathcal{S} \to \mathcal{A}$ a policy.
Let $P_\pi : \mathcal{S} \times \mathcal{S} \to [0,1]$ be the transition probability such that
$P_\pi(t,t')$ is the probability of transitioning from tail $t$ to tail $t'$
when the adversary picks the option that maximizes $\propose(P) + \nnext(N_P) +
	v(\tail(E_P))$. Let $R_\pi : \mathcal{S} \to \R$ also be the expected reward of
taking the action given by $\pi$ in the game. Lastly, let $\OPT(\cdot)$ be the
optimal reward of either the RANDAO MDP/game.
Using a similar argument to \cite{AW24}, we know that we can reduce the state
space to just the tail length by working with the RANDAO MDP.
\begin{lemmarep}
	\(\OPT(G) = \OPT(M)\) where $G$ is the RANDAO manipulation game and $M$ is
	the RANDAO MDP.
\end{lemmarep}
\begin{proof}
	We first observe that the game is ergodic, since each epoch has a non-zero
	probability of having samples with the same fixed epoch (or tail length).
	Hence, the optimal average reward is independent of the starting state.

	Now consider the RANDAO game $G$. Note that every epoch has a unique tail
	length. For any strategy playing the game, there is a
	corresponding strategy for the MDP that induces the same sequence of tails.
	As the game is ergodic, each state pair will be visited infinitely often,
	and in expectation these strategies will have the same average reward.
	Similarly, any strategy playing the MDP has a corresponding strategy in
	the game as the actions are the same. Hence, the optimal average reward
	must be the same.
\end{proof}

As discussed at the start of this section, computing the
transition probabilities for non-uniform rewards in closed form is not always
tractable.
We therefore propose sampling the transition probabilities to estimate them.
While sampling transitions from large tails is expensive, we observe that
we can truncate the MDP to small tails. Intuitively, as large tails are rare,
ignoring them will not change the optimal reward by much.

\begin{box1}[]{}
	\noindent\textbf{Approximate RANDAO Solver.}
	\begin{enumerate}
		\item Truncate the MDP to a state space $S^T = \{ 0, 1, \dots, T\}$ such
		      that all transitions that go to a tail $t > T$ now transition to $T$. The
		      reward of the transition is either $X$ for upper truncation or $0$ for lower
		      truncation. We call this the $T$-truncation of the model\footnote{We
			      take $X = R_{\max} \cdot \E[L(T,\alpha)]$, an upper bound on the expected
			      reward accrued during a large-tail sequence (see
			      Lemma~\ref{lem:large-tail-length}). As we show later, the upper and lower
			      truncations give upper and lower bounds.}.
		\item Initialize values $v(\cdot) = 0$ for each state.
		\item Empirically estimate the truncated transition probabilities and reward
		      vector with $N$ samples for each $P(t, \cdot)$ and $R(\cdot)$ to get matrix
		      $\hat{P}$ and vector $\hat{R}$.
		\item Compute average reward $\gamma$ for $\langle \hat{P}, \hat{R} \rangle$
		      using the stationary distribution of $\hat{P}$ and taking the product with
		      $\hat{R}$.
		\item Perform a policy improvement step by solving for values using $V +
			      \gamma = \hat{R} + \hat{P} V$ where $V$ is the value vector, and set
		      $v := V$. If values did not change more than the tolerance, output the
		      current valuation. Otherwise continue from step 3.
	\end{enumerate}
\end{box1}

\subsection{Reward Guarantees}
We will now analyze the optimality loss of \emph{truncation} and using the
\emph{approximate} model generated by samples. These bounds will ensure that we
truncate conservatively and sample enough times to accurately approximate the
optimal reward. Our analysis will depend on an upper bound on the maximum reward
of a single transition $R_{\max}$. For example, if the reward is the number of
blocks proposed, then $R_{\max} = 32$.

\subparagraph*{Truncation bound.}
Intuitively, we observe that any transition to a large tail $t > T$ must at
some point transition back to a small tail $t \leq T$ for bounded $\alpha$. Hence,
we can bound the expected number of large tail transitions and assign the
maximum reward $R_{\max}$ to each of these transitions to get an upper bound.
To do this, we first need to bound the expected length of consecutive large
tail transitions.

\begin{lemmarep} \label{lem:large-tail-length}
	For any policy $\pi$, let $L^\pi(T,\alpha)$ denote the number
	of consecutive rounds the adversary spends in tails greater than $T$,
	starting from $t = 32$. Let $L(T,\alpha)$ be the corresponding quantity
	under the policy that always picks the longest tail. Then $L(T,\alpha)$
	stochastically dominates $L^\pi(T,\alpha)$ for every $\pi$, and
	\[
		\E[L(T,\alpha)] \leq X_{T,\alpha}
	\]
	where $X_{T,\alpha}$ is the last element of the vector $(I - P_T)^{-1}
		\mathbf{1}$ where $P_T$ is the transition matrix of the Markov chain with
	state space $\{T+1, \dots, 32\}$ and transition probabilities of the
	strategy that always picks the longest tail.
\end{lemmarep}
\begin{proof}
	We start by observing that the strategy that always picks the longest tail
	stochastically dominates any other strategy in terms of the sequence of
	rounds spent in tails greater than $T$. Consider the next epoch
	samples across strategies: the tail-maximizing choice
	realizes a next-state tail length that is at least as
	large as any other strategy's. Moreover, choosing the max tail dominates
	returning to smaller states. Hence, the expected length of
	consecutive rounds in large tails of the tail-maximizing strategy is an
	upper bound on the expected length of consecutive rounds in large tails of
	any strategy.
	Lastly, we observe that the length of this path is maximized by starting
	from the largest tail $32$.
	Hence, as the expected length of rounds in large tails
	starting from each large state is given by the vector $(I - P_T)^{-1}
		\mathbf{1}$, the expected length starting from $32$ is the last element of
	this vector.
\end{proof}
\begin{figure}
    \centering
    \begin{tikzpicture}
        \begin{axis}[
            paperplot, percent x,
            xlabel={$\alpha$},
            ylabel={$\E[L(T,\alpha)]$},
            xmin=0, xmax=0.35,
            ymin=0.5, ymax=6,
            legend pos=north west,
        ]
        \addplot[color=modelPerSlot, smooth] coordinates {
            (0.0, 1.0) (0.01, 1.3491620120291277) (0.02, 1.9999957160858743)
            (0.03, 1.9999983321273445) (0.04, 2.0000045384805594) (0.05, 2.0000376389170658)
            (0.06, 2.0001512125147265) (0.07, 2.000486383154926) (0.08, 2.0012951992421835)
            (0.09, 2.0032497370361204) (0.1, 2.0073770659024106) (0.11, 2.016076143756933)
            (0.12, 2.03328659720166) (0.13, 2.065039922583126) (0.14, 2.122421355674821)
            (0.15, 2.2148760016777347) (0.16, 2.3522779025541705) (0.17, 2.5303892528777534)
            (0.18, 2.7170013317994632) (0.19, 2.8770219554931282) (0.2, 2.9776715136660217)
            (0.21, 3.026785655086704) (0.22, 3.0595789713557453) (0.23, 3.0996495019568813)
            (0.24, 3.1607566645610445) (0.25, 3.251779564612641) (0.26, 3.3795771823162224)
            (0.27, 3.5447364204461485) (0.28, 3.7372782762190386) (0.29, 3.9387658510019667)
            (0.3, 4.131850848034565) (0.31, 4.318395105377139) (0.32, 4.519539833370082)
            (0.33, 4.760266997389667) (0.34, 5.049036133289421) (0.35, 5.380043893151904)
            (0.36, 5.750408359796422) (0.37, 6.1733790893672476) (0.38, 6.671882745588231)
            (0.39, 7.267548635335706) (0.4, 7.986922143043298) (0.41, 8.874429033447393)
            (0.42, 9.998439583661282) (0.43, 11.467659993886507) (0.44, 13.470443017300163)
        };
        \addlegendentry{$T=5$}
        \addplot[color=modelUniform] coordinates {
            (0.0, 1.0) (0.01, 1.0042851774995194) (0.02, 1.240334058931573)
            (0.03, 1.9563241528564022) (0.04, 1.9999946654933787) (0.05, 1.9999937281025364)
            (0.06, 2.000000557146219) (0.07, 2.0000282898487867) (0.08, 2.0001036859410677)
            (0.09, 2.0002929926155506) (0.1, 2.000740841004676) (0.11, 2.0017855232353794)
            (0.12, 2.0040754890604893) (0.13, 2.008814156595942) (0.14, 2.0185734474284067)
            (0.15, 2.0372666522691683) (0.16, 2.0721901719071356) (0.17, 2.1337172842087666)
            (0.18, 2.232751137786132) (0.19, 2.3787344169626286) (0.2, 2.5612847539872696)
            (0.21, 2.749869616777619) (0.22, 2.902429511190607) (0.23, 2.991410658490218)
            (0.24, 3.037428680319448) (0.25, 3.0749077218447782) (0.26, 3.1284028107677297)
            (0.27, 3.212434241424) (0.28, 3.336907953257839) (0.29, 3.505649158357378)
            (0.3, 3.7094153131508683) (0.31, 3.92686299139126) (0.32, 4.137065212097266)
            (0.33, 4.343355156122458) (0.34, 4.573842553538165) (0.35, 4.856002055968187)
            (0.36, 5.1941142951614925) (0.37, 5.581435314364408) (0.38, 6.026074254327437)
            (0.39, 6.55514546821128) (0.4, 7.198609987009984) (0.41, 7.992337069658251)
            (0.42, 8.996630657468405) (0.43, 10.309449685064516) (0.44, 12.098360284709997)
        };
        \addlegendentry{$T=6$}
        \addplot[color=modelThreshold] coordinates {
            (0.0, 1.0) (0.01, 1.0000429143210723) (0.02, 1.0054823539421485)
            (0.03, 1.0896543074456753) (0.04, 1.5052394498348605) (0.05, 1.965097988174568)
            (0.06, 1.9999855136515006) (0.07, 1.99999621537675) (0.08, 2.000008295259182)
            (0.09, 2.0000263731191277) (0.1, 2.000074113503384) (0.11, 2.000196608307241)
            (0.12, 2.000488163382396) (0.13, 2.001148579198712) (0.14, 2.0026323809174036)
            (0.15, 2.0057289744707485) (0.16, 2.012141206529745) (0.17, 2.024956908011784)
            (0.18, 2.0494105262091296) (0.19, 2.094556609230153) (0.2, 2.1717271546230994)
            (0.21, 2.29299335493901) (0.22, 2.461025689457132) (0.23, 2.6550443318431802)
            (0.24, 2.835071238128778) (0.25, 2.958246575930584) (0.26, 3.0235179062499347)
            (0.27, 3.066413434858988) (0.28, 3.12110752326973) (0.29, 3.209063944643565)
            (0.3, 3.3432263192618064) (0.31, 3.527936347303118) (0.32, 3.7505890053867854)
            (0.33, 3.9848875448716004) (0.34, 4.212234633494211) (0.35, 4.448143640174179)
            (0.36, 4.729894266051408) (0.37, 5.078001196529122) (0.38, 5.488627932012063)
            (0.39, 5.966807108241286) (0.4, 6.543418833098287) (0.41, 7.258734086813231)
            (0.42, 8.163774782760715) (0.43, 9.345906627418069) (0.44, 10.956302701319853)
        };
        \addlegendentry{$T=7$}
        \end{axis}
    \end{tikzpicture}
    \caption{Bound on $\E[L(T,\alpha)]$ from Lemma~\ref{lem:large-tail-length},
    plotted versus stake fraction $\alpha$ for truncation thresholds
    $T \in \{5,6,7\}$.}
    \label{fig:tail-length}
\end{figure}
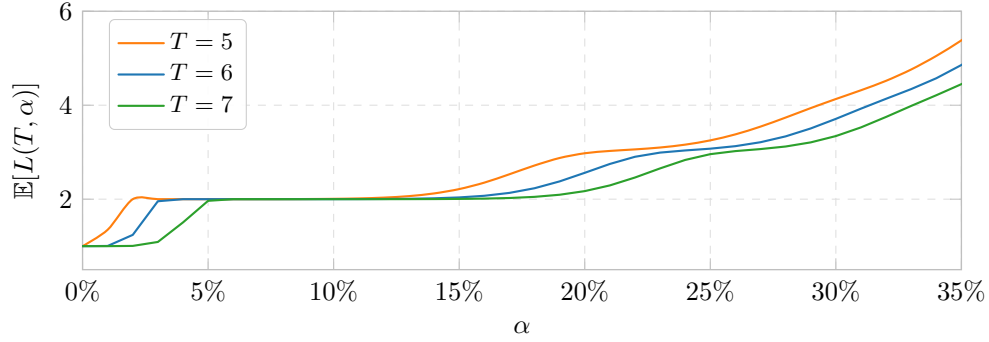
 \begin{theoremrep}
	Let $M^+$ and $M^-$ be the upper and lower $T$-truncations of the RANDAO MDP
	$M$. Then,
	\(\OPT(M^-) \leq \OPT(M) \leq \OPT(M^+)\)
\end{theoremrep}
\begin{proof}
	$M^-$ differs from $M$ only by replacing the
	reward on each transition to a large tail with $0$, so any policy's
	average reward in $M^-$ is at most its average reward in $M$. Hence, $\OPT(M^-) \leq \OPT(M)$.
	Now consider the optimal
	policy $\pi^*$ for $M$. Observe that whenever $\pi^*$ transitions to a large
	tail $t > T$, it must eventually transition back to a small tail $t \leq T$.
	As this will last less than $L(T,\alpha)$ rounds in expectation, we can
	upper bound the expected reward of this path by $R_{\max} \cdot
		\E[L(T,\alpha)]$. Therefore, the corresponding policy in $M^+$ that takes
	the same actions as $\pi^*$ in small tails and transitions to $T$ with
	reward $R_{\max} \cdot \E[L(T,\alpha)]$ in large tails will have an expected
	reward at least as much as $\pi^*$. Moreover, since the $M^+$
	policy collapses each large-tail sequence into a single transition while
	$\pi^*$ takes $\E[L(T,\alpha)]$ steps in expectation, the $M^+$ policy
	accumulates at least as much reward over strictly fewer transitions, so its
	long-run average reward is at least that of $\pi^*$. Hence,
	$\OPT(M) \leq \OPT(M^+)$ as desired.
\end{proof}

\subparagraph*{Sampling complexity.}
We now observe that by an application of the Chernoff-Hoeffding bound and a
union bound over the $(T{+}1)^2$ entries of the transition matrix and the
$(T{+}1)$ entries of the reward vector, we can derive the
sufficient number of samples to estimate the probabilities and rewards up to an
error of $\varepsilon$ with probability $1 - \delta$.
\begin{lemmarep}
	Given $\varepsilon, \delta > 0$, let $\hat{P}$ and $\hat{R}$ be the
	empirical estimates of the transition probabilities and rewards using $N$
	independent samples per state from the truncated MDP $M^T$ with a fixed policy
	where
	\[
		N \geq \max\left(
		\frac{(T+1)^2}{2\varepsilon^2} \ln\left(\frac{4(T+1)^2}{\delta}\right),
		\frac{R_{\max}^2}{2\varepsilon^2} \ln\left(\frac{4(T+1)}{\delta}\right)
		\right)
	\]
	Then, with probability at least $1 - \delta$, both of the following hold:
	\begin{align*}
		 & \sum_{t'} | P(t,t') - \hat{P}(t,t') | \leq \varepsilon,
		\quad \forall t \in \{0,\dots,T\}                          \\
		 & | R(t) - \hat{R}(t) | \leq \varepsilon,
		\quad \forall t \in \{0,\dots,T\}
	\end{align*}
\end{lemmarep}
\begin{proof}
	We proceed with applying a Chernoff-Hoeffding bound on each estimate
	and then taking a union bound over all the estimates.
	For each state $t$, we have $N$ independent samples of the next state
	$t'$. Let $X_i$ be the indicator random variable for the event that
	the $i\textsuperscript{th}$ sample transitioned to $t'$.
	Then, $\hat{P}(t,t') = \frac{1}{N} \sum_{i=1}^{N} X_i$
	and $\E[\hat{P}(t,t')] = P(t,t')$. By the Chernoff-Hoeffding bound, we have
	\[
		\Pr\left[ | \hat{P}(t,t') - P(t,t') | \geq \frac{\varepsilon}{T+1} \right]
		\leq 2\exp\left(-\frac{2N\varepsilon^2}{(T+1)^2}\right)
	\]
	Now, we know that $\sum_{t'} | P(t,t') - \hat{P}(t,t') | \leq
		\varepsilon$ if $| P(t,t') - \hat{P}(t,t') | \leq \varepsilon/(T+1)$ for all $t'$.
	Therefore, by a union bound over all $t'$, we have
	\[
		\Pr\left[ \bigcup_{t'} \left[
				| \hat{P}(t,t') - P(t,t') | \geq \frac{\varepsilon}{T+1}
				\right] \right]
		\leq 2(T+1)\exp\left(-\frac{2N\varepsilon^2}{(T+1)^2}\right)
	\]
	Hence, by another union bound over all $t$, we have
	\[
		\sum_{t'} | P(t,t') - \hat{P}(t,t') |
		\leq
		\varepsilon, \quad \forall t \in \{0,\dots,T\}
	\]
	with probability at least $1 - 2 (T+1)^2 \exp(-2N\varepsilon^2/(T+1)^2)$.
	As $N \geq \frac{(T+1)^2}{2\varepsilon^2}
		\allowbreak
		\ln\left(\frac{4(T+1)^2}{\delta}\right)$,
	this is $\geq 1 - \delta/2$.

	Now consider the reward estimates. Let $Y_i$ be the random variable
	representing the reward of the $i\textsuperscript{th}$ sample from state
	$t$. Then, $\hat{R}(t) = \frac{1}{N} \sum_{i=1}^{N} Y_i$ and
	$\E[\hat{R}(t)] = R(t)$. By the Chernoff-Hoeffding bound,
	using the fact that $Y_i \in [0, R_{\max}]$, we have
	\[
		\Pr\left[ | \hat{R}(t) - R(t) | \geq \varepsilon \right]
		\leq 2\exp\left(-\frac{2N\varepsilon^2}{R_{\max}^2}\right)
	\]
	Since there are $T+1$ states, by a union bound over all $t$, we have
	\[
		| R(t) - \hat{R}(t) | \leq \varepsilon, \quad \forall t \in \{0,\dots,T\}
	\]
	with probability at least $1 - 2 (T+1) \exp(-2N\varepsilon^2/R_{\max}^2)$. As
	$N \geq \frac{R_{\max}^2}{2\varepsilon^2}
		\ln\left(\frac{4(T+1)}{\delta}\right)$, this is $\geq 1 - \delta/2$.

	Therefore, by a final union bound, we have that
	both properties hold with probability at least $1 - \delta$ as desired.
\end{proof}
For $\varepsilon = 0.0085$ and $\delta = 0.01$,
$N = 10^8$ samples per state is sufficient for $T = 7$ and the $R_{\max}$
of the reward models we consider.
We will be using this setting in our experiments.
\section{MEV Models and Results} \label{sec:results}

In practice, MEV is a complex phenomenon that is affected by not just the number
of blocks, but their successive combinations or the result of a stochastic
process. We will therefore consider a few different MEV models that capture some
of these phenomena.
For all reported utilities, we plot the reward per slot (i.e.\ the average reward
for a single slot) against $\alpha$.
We focus on stake fractions $\alpha \leq 0.35$ throughout: at
larger $\alpha$ the expected time spent in large tails $\E[L(T,\alpha)]$
(Lemma~\ref{lem:large-tail-length}, \cref{fig:tail-length}) grows rapidly, and
the truncation bound $R_{\max} \cdot \E[L(T,\alpha)]$ separating our
upper and lower estimates degrades.

\subsection{Warm-up: Reproducing Uniform Reward Results} \label{subsec:warmup} To
validate our methodology, we first reproduce the results of \cite{AW24} in the
case of uniform rewards. In this case, the reward model $\mathcal{R}$ is defined
as the number of 1's in the proposal string for both $\propose$ and $\nnext$.
The results are shown in \cref{tab:uniform}.

\begin{table}[H]
    \centering
    \caption{Comparison of the optimal rewards reported by \cite{AW24}
    with our method's $U$ (upper truncation) and $L$ (lower truncation)
    bounds for uniform rewards (a single unit per proposed block). Each
    entry has an error of at most $0.0265\%$ ($\varepsilon / 32$).}
    \label{tab:uniform}
    \setlength{\tabcolsep}{4pt}
    \begin{tabular}{l|rrrrrrr}
    \hline
    \multicolumn{1}{c|}{$\alpha$} & $5\%$ & $10\%$ & $15\%$ & $20\%$ & $25\%$ & $30\%$ & $35\%$ \\
    \cline{2-8}
    \hline
    \cite{AW24} & 5.0483\% & 10.1881\% & 15.3996\% & 20.6777\% & 26.0247\% & 31.4516\% & 36.9734\% \\
    $U$ & 5.0483\% & 10.1879\% & 15.3998\% & 20.6783\% & 26.0323\% & 31.4916\% & 37.2290\% \\
    $L$ & 5.0483\% & 10.1878\% & 15.3997\% & 20.6776\% & 26.0250\% & 31.4487\% & 36.9581\% \\
    \hline
    \end{tabular}
\end{table}
 
\subsection{Per-slot MEV}
We define the per-slot MEV reward model as follows (Example~\ref{ex:perslot}). In every slot,
there is a base reward of 1 in two components:
\begin{itemize}
	\item \emph{Ephemeral reward.} This is the reward for proposing a block in
	      that slot. If the block is not proposed, this reward is lost.
	\item \emph{Rollover reward.} This is the reward that rolls over from the
	      previous slot if it was not proposed. If the block is proposed, this reward
	      is claimed. If the block is not proposed, this reward rolls over to the
	      next slot.
\end{itemize}
Intuitively, there are several reasons why this model is
reasonable. Certain consensus-layer block rewards are ephemeral and are
lost if the current block is not proposed, while many execution-layer
rewards (such as priority fees and arbitrage or liquidation
opportunities) roll over to the next block if not claimed. Even some of
the consensus rewards on Ethereum roll over: the block immediately
following a missed slot contains approximately twice the proposer
reward (see \cref{fig:cl_rewards}) due to the additional attestations
that can be included.
We will use 1 reward per slot with $60\%$ ephemeral and $40\%$ rollover
reward. Let $\nnext(n_t)$ be defined simply as the number of 1s in $p_{t+1}$.
Let $\propose(p_t)$ be defined such that each proposed block gives a reward of
$1.0 + r$, where $r$ is the amount of rollover reward from previous slots.
Each missed slot accumulates $0.4$ rollover, and
a proposal claims and resets it to $0$.
Using $R_{\max} = 32$ for this model, we run our approximate RANDAO
solver and the results are shown in \cref{tab:per-slot}.
\begin{table}[H]
    \centering
    \caption{Optimal rewards for the per-slot MEV reward model.
    $U$ (upper truncation) and $L$ (lower truncation) are compared
    with the honest strategy $H$. Each entry has an error of at most
    $0.0265\%$ ($\varepsilon / 32$).}
    \label{tab:per-slot}
    \setlength{\tabcolsep}{4pt}
    \begin{tabular}{l|rrrrrrr}
    \hline
    \multicolumn{1}{c|}{$\alpha$} & $5\%$ & $10\%$ & $15\%$ & $20\%$ & $25\%$ & $30\%$ & $35\%$ \\
    \cline{2-8}
    \hline
    $U$ & 5.0493\% & 10.1922\% & 15.4107\% & 20.7007\% & 26.0734\% & 31.5628\% & 37.3514\% \\
    $L$ & 5.0493\% & 10.1922\% & 15.4106\% & 20.6999\% & 26.0660\% & 31.5194\% & 37.0754\% \\
    $H$ & 4.9999\% & 10.0009\% & 15.0015\% & 20.0017\% & 25.0016\% & 30.0018\% & 35.0016\% \\
    \hline
    \end{tabular}
\end{table}
\subsection{Multi-block MEV}
We now define the multi-block MEV reward models as follows.
In every epoch, the rewards scale with the number of consecutive blocks
proposed. We will consider two variants of this model:
\begin{itemize}
	\item \textbf{Smooth Model.} The reward for proposing $k$ consecutive
	      blocks is $k\sqrt{k}$, a smooth superlinear shape that captures
	      phenomena such as arbitrage across multiple blocks
	      (Example~\ref{ex:smooth}).
	\item \textbf{Threshold Model.} If $k$ consecutive blocks are proposed,
	      the reward is $k$ for $k < 5$ and $k + 10$ for $k \geq 5$. This
	      captures phenomena such as oracle manipulation, where a proposer
	      with a long enough run can manipulate an oracle to their advantage
	      and gain a large reward (Example~\ref{ex:threshold}).
\end{itemize}
Both multi-block reward models above are stylized illustrations rather
than calibrated estimates. The $k^{3/2}$ shape is loosely inspired by
CFAMM arbitrage with fees~\cite{milionis2024automated}: fees create a
no-trade band that the cross-venue price gap must cross before
arbitrage becomes profitable, and in the regime relevant to short
multi-block sequences the resulting option-like payoff can grow
superlinearly in $k$. The threshold choice ($k = 5$, bonus $10$) is
similarly illustrative. Documented TWAP attacks have their sharp
discontinuity at $k = 2$~\cite{mackinga2022twap}. Our purpose is to
showcase a methodology that can quantify the impact of any such
reward structure, should one emerge in practice for multi-block
strategies that are not yet exploited at scale.
The results for both models are shown below in
\cref{tab:multi-block-vertical}.
\begin{table}[H]
    \centering
    \caption{Optimal rewards for multi-block MEV models. Results for the smooth reward model (top) and threshold reward model (bottom) are shown.
        $U$ (upper truncation) and $L$ (lower truncation) are compared with the honest strategy $H$.}
    \label{tab:multi-block-vertical}
    \setlength{\tabcolsep}{4pt}
    \begin{tabular}{l|rrrrrrr}
    \hline
    \multicolumn{8}{c}{\textbf{Smooth Rewards}} \\[0.5ex]
    \hline
    \multicolumn{1}{c|}{$\alpha$} & $5\%$ & $10\%$ & $15\%$ & $20\%$ & $25\%$ & $30\%$ & $35\%$ \\
    \cline{2-8}
    \hline
    $U$ & 5.2616\% & 11.0913\% & 17.5533\% & 24.7437\% & 32.8185\% & 42.0835\% & 53.7479\% \\
    $L$ & 5.2616\% & 11.0913\% & 17.5529\% & 24.7392\% & 32.7769\% & 41.8354\% & 52.1490\% \\
    $H$ & 5.2072\% & 10.8585\% & 17.0013\% & 23.6931\% & 31.0040\% & 39.0198\% & 47.8438\% \\
    \hline
    \multicolumn{8}{c}{} \\[1ex]
    \hline
    \multicolumn{8}{c}{\textbf{Threshold Rewards}} \\[0.5ex]
    \hline
    \multicolumn{1}{c|}{$\alpha$} & $5\%$ & $10\%$ & $15\%$ & $20\%$ & $25\%$ & $30\%$ & $35\%$ \\
    \cline{2-8}
    \hline
    $U$ & 5.0539\% & 10.2793\% & 15.8691\% & 22.1828\% & 29.7048\% & 38.9595\% & 50.7152\% \\
    $L$ & 5.0539\% & 10.2793\% & 15.8689\% & 22.1806\% & 29.6848\% & 38.8404\% & 49.9715\% \\
    $H$ & 5.0052\% & 10.0829\% & 15.3942\% & 21.1721\% & 27.6875\% & 35.2180\% & 44.0074\% \\
    \hline
    \end{tabular}
\end{table}
\section{Improvement over Honest Strategy} \label{sec:improvement}
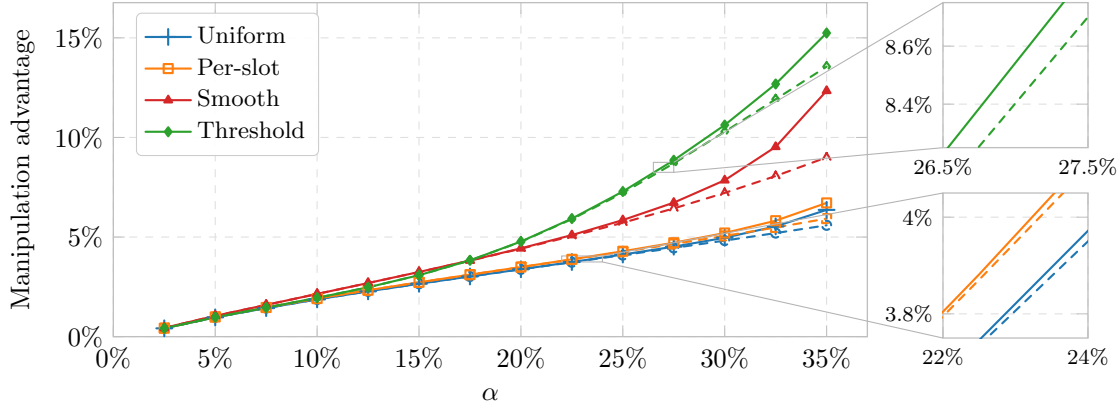
\begin{figure}[h]
	\centering
	\begin{tikzpicture}
		\begin{axis}[
				name=main,
				paperplot, percent x, percent y,
				unbounded coords=jump,
				width=0.7\linewidth, height=6cm,
				xlabel={$\alpha$},
				ylabel={Manipulation advantage},
				xmin=0, xmax=0.37,
				ymin=0,
				legend pos=north west,
			]
			\addplot[color=modelUniform,   mark=+, mark size=3.5pt, thick]  table[x=alpha, y=U_uniform]   {figures/data/improvement.dat};
			\addplot[color=modelUniform,   mark=o, dashed, forget plot]     table[x=alpha, y=L_uniform]   {figures/data/improvement.dat};
			\addplot[color=modelPerSlot,   mark=square]                     table[x=alpha, y=U_perslot]   {figures/data/improvement.dat};
			\addplot[color=modelPerSlot,   mark=square, dashed, forget plot] table[x=alpha, y=L_perslot]   {figures/data/improvement.dat};
			\addplot[color=modelSmooth,    mark=triangle*] table[x=alpha, y=U_smooth]   {figures/data/improvement.dat};
			\addplot[color=modelSmooth,    mark=triangle, dashed, forget plot] table[x=alpha, y=L_smooth]   {figures/data/improvement.dat};
			\addplot[color=modelThreshold, mark=diamond*] table[x=alpha, y=U_threshold] {figures/data/improvement.dat};
			\addplot[color=modelThreshold, mark=diamond, dashed, forget plot] table[x=alpha, y=L_threshold] {figures/data/improvement.dat};
			\legend{Uniform, Per-slot, Smooth, Threshold}
			\draw[gray!60, thin] (axis cs:0.22,3.75) rectangle (axis cs:0.24,4.05);
			\coordinate (lowbox-tr) at (axis cs:0.24,4.05);
			\coordinate (lowbox-br) at (axis cs:0.24,3.75);
			\draw[gray!60, thin] (axis cs:0.265,8.25) rectangle (axis cs:0.275,8.75);
			\coordinate (highbox-tr) at (axis cs:0.275,8.75);
			\coordinate (highbox-br) at (axis cs:0.275,8.25);
		\end{axis}

		\begin{axis}[
				name=zoomHigh,
				paperplot,
				at={(main.north east)}, anchor=north west,
				xshift=1.0cm,
				width=3.5cm, height=3.5cm,
				xmin=0.265, xmax=0.275,
				ymin=8.25, ymax=8.75,
				xtick={0.265, 0.275},
				ytick={8.4, 8.6},
				xticklabel={\pgfmathparse{\tick*100}\pgfmathprintnumber[fixed, precision=1]{\pgfmathresult}\%},
				yticklabel={\pgfmathprintnumber[fixed, precision=1]{\tick}\%},
				tick label style={font=\scriptsize},
				label style={font=\scriptsize},
				axis background/.style={fill=white},
			]
			\addplot[color=modelUniform]   table[x=alpha, y=U_uniform]   {figures/data/improvement.dat};
			\addplot[color=modelUniform,   dashed] table[x=alpha, y=L_uniform]   {figures/data/improvement.dat};
			\addplot[color=modelPerSlot]   table[x=alpha, y=U_perslot]   {figures/data/improvement.dat};
			\addplot[color=modelPerSlot,   dashed] table[x=alpha, y=L_perslot]   {figures/data/improvement.dat};
			\addplot[color=modelSmooth]    table[x=alpha, y=U_smooth]   {figures/data/improvement.dat};
			\addplot[color=modelSmooth,    dashed] table[x=alpha, y=L_smooth]   {figures/data/improvement.dat};
			\addplot[color=modelThreshold] table[x=alpha, y=U_threshold] {figures/data/improvement.dat};
			\addplot[color=modelThreshold, dashed] table[x=alpha, y=L_threshold] {figures/data/improvement.dat};
		\end{axis}

		\begin{axis}[
				name=zoomLow,
				paperplot,
				at={(zoomHigh.south west)}, anchor=north west,
				yshift=-0.6cm,
				width=3.5cm, height=3.5cm,
				xmin=0.22, xmax=0.24,
				ymin=3.75, ymax=4.05,
				xtick={0.22, 0.24},
				ytick={3.8, 4.0},
				xticklabel={\pgfmathparse{\tick*100}\pgfmathprintnumber[fixed, precision=0]{\pgfmathresult}\%},
				yticklabel={\pgfmathprintnumber[fixed, precision=1]{\tick}\%},
				tick label style={font=\scriptsize},
				label style={font=\scriptsize},
				axis background/.style={fill=white},
			]
			\addplot[color=modelUniform]   table[x=alpha, y=U_uniform]   {figures/data/improvement.dat};
			\addplot[color=modelUniform,   dashed] table[x=alpha, y=L_uniform]   {figures/data/improvement.dat};
			\addplot[color=modelPerSlot]   table[x=alpha, y=U_perslot]   {figures/data/improvement.dat};
			\addplot[color=modelPerSlot,   dashed] table[x=alpha, y=L_perslot]   {figures/data/improvement.dat};
			\addplot[color=modelSmooth]    table[x=alpha, y=U_smooth]   {figures/data/improvement.dat};
			\addplot[color=modelSmooth,    dashed] table[x=alpha, y=L_smooth]   {figures/data/improvement.dat};
			\addplot[color=modelThreshold] table[x=alpha, y=U_threshold] {figures/data/improvement.dat};
			\addplot[color=modelThreshold, dashed] table[x=alpha, y=L_threshold] {figures/data/improvement.dat};
		\end{axis}

		\draw[gray!60, thin] (highbox-tr) -- (zoomHigh.north west);
		\draw[gray!60, thin] (highbox-br) -- (zoomHigh.south west);
		\draw[gray!60, thin] (lowbox-tr)  -- (zoomLow.north west);
		\draw[gray!60, thin] (lowbox-br)  -- (zoomLow.south west);
	\end{tikzpicture}
	\caption{Manipulation advantage over the honest strategy across
		reward models, measured as $(\OPT(M^\mathcal{R}) -
			h^\mathcal{R})/h^\mathcal{R}$. Solid lines with filled markers show
		the upper-truncation bound. Dashed lines show
		the lower-truncation bound.}
	\label{fig:improvement}
\end{figure}

The severity of the incentive to manipulate RANDAO can be measured by the
percentage increase in expected rewards per time over the honest strategy
of proposing in every slot. We use this metric to compare the effects of
different reward models on RANDAO manipulation incentives.
\begin{definition}[Improvement]
	Let $M^\mathcal{R}$ be a RANDAO MDP with reward model $\mathcal{R}$
	and $h^\mathcal{R}$ be the average reward of the honest strategy
	in $M^\mathcal{R}$. Then, the improvement over the honest strategy is
	defined as:
	\(
	\frac{\OPT(M^\mathcal{R}) - h^\mathcal{R}}{h^\mathcal{R}}
	\)
\end{definition}
We find that this incentive is highly dependent on the reward model used
as shown in \cref{fig:improvement}. This suggests that as the MEV extraction
landscape evolves, the incentive to manipulate RANDAO may also change.

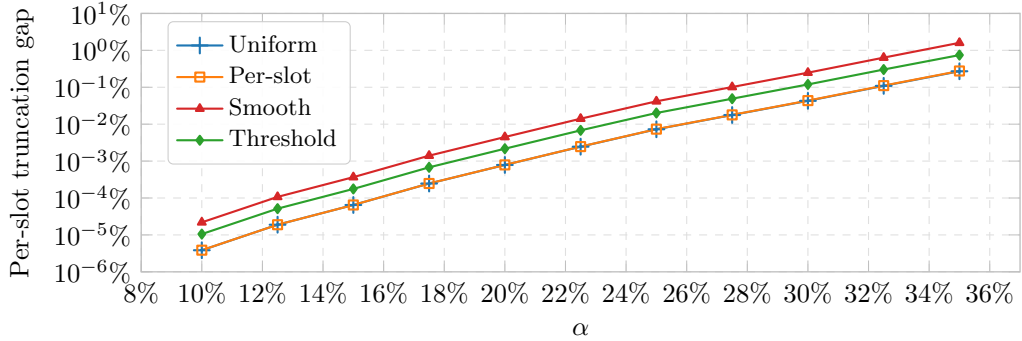
\begin{figure}[h]
    \centering
    \begin{tikzpicture}
        \begin{axis}[
            paperplot, percent x,
            unbounded coords=jump,
            restrict x to domain=0.1:1,
            xlabel={$\alpha$},
            ylabel={Per-slot truncation gap},
            ymode=log,
            log basis y=10,
            yticklabel={$10^{\pgfmathprintnumber[fixed, precision=0]{\tick}}\%$},
            ytick={1e-6,1e-5,1e-4,1e-3,1e-2,1e-1,1,10},
            minor y tick num=9,
            xmin=0.08, xmax=0.37,
            ymin=1e-6, ymax=10,
            legend pos=north west,
        ]
        \addplot[color=modelUniform,   mark=+, mark size=3.5pt, thick]      table[x=alpha, y=uniform]   {figures/data/error.dat};
        \addplot[color=modelPerSlot,   mark=square]                          table[x=alpha, y=perslot]   {figures/data/error.dat};
        \addplot[color=modelSmooth,    mark=triangle*] table[x=alpha, y=smooth]    {figures/data/error.dat};
        \addplot[color=modelThreshold, mark=diamond*]  table[x=alpha, y=threshold] {figures/data/error.dat};
        \legend{Uniform, Per-slot, Smooth, Threshold}
        \end{axis}
    \end{tikzpicture}
    \caption{Per-slot gap between upper- and lower-truncation estimates of
    the optimal average reward at truncation tail length $T = 7$, across
    reward models and stake fractions $\alpha$.}
    \label{fig:error}
\end{figure}

\subparagraph*{Error Analysis.}
We plot the absolute error between the upper and lower truncation in
\cref{fig:error}. The error is small for small values of $\alpha$, as
expected, and the primary source of error is sampling.
For larger values of $\alpha$, the error increases as the optimal policy
starts to depend more on the large tail states. However,
even for $\alpha = 0.35$, the difference remains small.
If higher precision is required, more samples and higher truncation $T$ can be used.

\begin{figure}[h]
	\centering
	\begin{tikzpicture}
		\begin{axis}[
				paperplot, percent y,
				xlabel={Epoch length $\ell$},
				ylabel={Manipulation advantage},
				xmin=12, xmax=140,
				ymin=0,
				legend pos=north east,
				xtick={16,32,64,96,128},
				extra x ticks={32},
				extra x tick labels={},
				extra x tick style={grid=major, grid style={dashed, red!60, thick}},
			]
			\addplot[color=modelUniform,   mark=+, mark size=3.5pt, thick]  table[x=L, y=uniform_035]   {figures/data/epoch.dat};
			\addplot[color=modelPerSlot,   mark=square]                     table[x=L, y=perslot_035]   {figures/data/epoch.dat};
			\addplot[color=modelSmooth,    mark=triangle*] table[x=L, y=smooth_035]    {figures/data/epoch.dat};
			\addplot[color=modelThreshold, mark=diamond*]  table[x=L, y=threshold_035] {figures/data/epoch.dat};
			\legend{Uniform, Per-slot, Smooth, Threshold}
		\end{axis}
	\end{tikzpicture}
	\caption{Relative manipulation gain $(\OPT(M^\mathcal{R}_\ell) -
			h^\mathcal{R}_\ell)/h^\mathcal{R}_\ell$ versus epoch length $\ell$
		at $\alpha = 0.35$, one curve per MEV reward model. The vertical
		red dashed line at $\ell = 32$ marks the current Ethereum value.}
	\label{fig:epoch_sweep}
\end{figure}
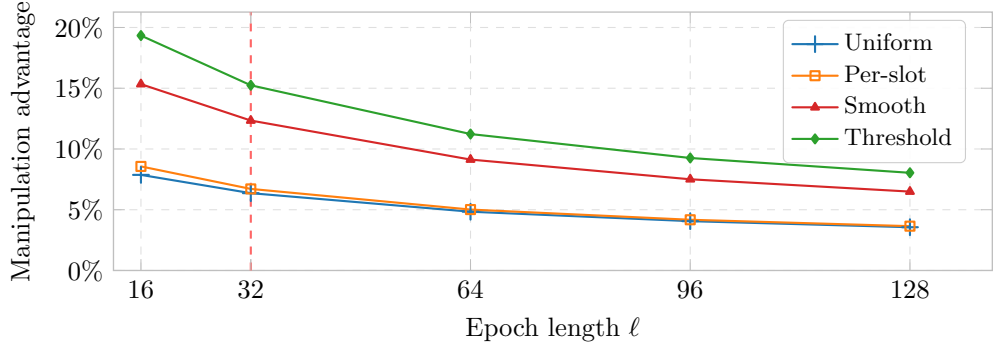

\subparagraph*{Epoch-Length Sensitivity.}
A natural protocol parameter to revisit is the epoch length $\ell$ itself,
currently fixed at $\ell = 32$ slots on Ethereum mainnet. Let
$M^\mathcal{R}_\ell$ denote the RANDAO MDP for reward model $\mathcal{R}$
under epoch length $\ell$, and $h^\mathcal{R}_\ell$ the corresponding
honest-strategy average reward.
\cref{fig:epoch_sweep} reports the relative manipulation gain
$(\OPT(M^\mathcal{R}_\ell) - h^\mathcal{R}_\ell) / h^\mathcal{R}_\ell$ as
we sweep $\ell$, holding $\alpha$ fixed and computed separately for each
of the four MEV reward models of \cref{sec:results}.
Across all four
models, we observe that the manipulation incentive decays with increased
epoch length.
Lengthening the epoch is therefore a partial mitigation, but for any
concrete $\ell$ the manipulation incentive depends on the
reward model.

\section{Tail Slashing} \label{sec:mitigation}

The preceding sections characterize how the incentive to manipulate
RANDAO depends on the reward model. A natural next question is whether
the reward structure itself can be modified to disincentivize
manipulation. Introducing a missed-slot penalty in normal
operation has been discussed within the Ethereum
community~\cite{missedslot_tempcheck}. In
this section we apply our methodology in reverse: rather than computing
the optimal adversarial strategy for a fixed reward model, we ask, for
each MEV reward model, what is a minimal slashing curve that we can
apply to the tail slots under which honest behavior becomes
$\varepsilon$-optimal? We model
tail slashing as a negative component of the reward
function and reuse the solver.

\subsection{Slashing as a Reward Model}

We now pick an illustrative slashing scheme that penalizes tail misses.
Let the epoch have length $\ell = 32$ and let $q \in \{0,1\}^t$ denote
the adversary's proposal pattern over a tail of length $t$, with
$q_i = 1$ if the slot at position $i$ from the end of the
epoch was proposed ($i = 0$ is the last slot). Manipulation is driven
by control of the final slots of an epoch, so we concentrate the
penalty there. The penalty is driven by two factors: how close the
missed slot is to the end of the epoch, and how many other misses
occur in the same tail.

\subparagraph*{Slashing scheme.}
Consider the following slashing scheme parameterized by a scaling constant
$\kappa$. For a fixed window $W=8$,
in every epoch, consider the last $W$ slots.
For each missed slot $i$ in the tail (numbered such that $i=0$ is the last
slot), we will assign a slashing penalty of $\kappa \cdot \frac{W-i}{W} \cdot D$
where $D$ is an additional multiplier equal to
$1 + (\text{\# of missed slots in the window})/W$.
We run our experiments and discuss manipulability for this particular
formula. For the game where every honest player proposes against a
single $\alpha$-adversary, we can write this as the following.

\begin{definition}[Tail slashing]
	\label{def:linear-slashing}
	Fix a window size $W \in \N$ and a slashing magnitude $\kappa \geq 0$.
	Index tail slots so that $i = 0$ is the last slot of the epoch. For a
	tail-proposal string $q$, define
	\[
		S(q) =
		\sum_{i=0}^{\min(t,W)-1} \mathbf{1}[q_i=0] \cdot \tfrac{W-i}{W},
		\qquad
		C(q) =
		\bigl|\{ i < \min(t,W) : q_i = 0 \}\bigr|.
	\]
	The slashing penalty for $q$ is
	\[
		s_\kappa(q) = -\kappa \cdot \Bigl(1 + \tfrac{C(q)}{W}\Bigr) \cdot S(q)
	\]
\end{definition}

\begin{example}[slashing with $W = 8$]
	\label{ex:slashing-effect}
	We list the penalty for several tail-proposal patterns. Each binary
	string shows the proposal pattern over the window, read so the
	rightmost bit is the last slot of the epoch ($i = 0$).
	\begin{alignat*}{4}
		 & \mathtt{11111110} \;\; &  & : \; s_\kappa = -\kappa \cdot \bigl(1 + \tfrac{1}{W}\bigr) \cdot 1 \;\;               &  & \approx -1.125\,\kappa \;\; &  & (\text{miss at the last slot})       \\
		 & \mathtt{01111111} \;\; &  & : \; s_\kappa = -\kappa \cdot \bigl(1 + \tfrac{1}{W}\bigr) \cdot \tfrac{1}{W} \;\;    &  & \approx -0.141\,\kappa \;\; &  & (\text{miss at } i = W{-}1)          \\
		 & \mathtt{11111100} \;\; &  & : \; s_\kappa = -\kappa \cdot \bigl(1 + \tfrac{2}{W}\bigr) \cdot \tfrac{2W-1}{W} \;\; &  & \approx -2.344\,\kappa \;\; &  & (\text{two clustered misses at end}) \\
		 & \mathtt{00000000} \;\; &  & : \; s_\kappa = -\kappa \cdot \bigl(1 + \tfrac{W}{W}\bigr) \cdot \tfrac{W+1}{2} \;\;  &  & = -9\,\kappa \;\;           &  & (\text{all window slots missed})
	\end{alignat*}
\end{example}

\subparagraph*{Slashed reward model.}
Given a reward model $\mathcal{R} = \langle \propose, \nnext \rangle$,
the slashed model is
\[
	\mathcal{R}^s_\kappa
	= \langle \propose + s_\kappa, \, \nnext \rangle.
\]
Slashing only modifies the $\propose$ component
not $\nnext$. We write $M^{\mathcal{R}^s_\kappa}$
for the RANDAO MDP (\cref{sec:mdp}) instantiated with
$\mathcal{R}^s_\kappa$.

\begin{definition}[Slashing threshold]
	\label{def:kappa-min}
	For reward model $\mathcal{R}$, stake fraction $\alpha$, and
	tolerance $\varepsilon > 0$, the slashing threshold
	$\kappa_{\min}$ is the minimum
	slashing magnitude $\kappa \geq 0$ for which honest behavior is
	$\varepsilon$-optimal in the slashed MDP
	$M^{\mathcal{R}^s_\kappa}$. That is, the optimal reward
	$\OPT(M^{\mathcal{R}^s_\kappa})$ is at most $\varepsilon$ above the
	honest average reward $h^\mathcal{R}$.
\end{definition}
The following monotonicity ensures that $\kappa_{\min}$ is well-defined.

\begin{lemmarep}[Monotonicity in $\kappa$] \label{lem:slashing-monotone}
	$\OPT(M^{\mathcal{R}^s_\kappa})$ is non-increasing in $\kappa$.
\end{lemmarep}
\begin{proof}
	Fix $\kappa < \kappa'$ and let $\pi$ be optimal for
	$M^{\mathcal{R}^s_{\kappa'}}$. On every transition where $\pi$ skips a
	tail slot, the slashing penalty under $\kappa'$ is at least as large
	as under $\kappa$. On transitions with no skip, the reward is identical.
	Since both MDPs are ergodic, the long-run average reward of $\pi$ is
	well-defined and starting-state independent in each. Hence the average
	reward of $\pi$ in $M^{\mathcal{R}^s_\kappa}$ is at
	least its average reward in $M^{\mathcal{R}^s_{\kappa'}}$, giving
	$\OPT(M^{\mathcal{R}^s_\kappa}) \geq \OPT(M^{\mathcal{R}^s_{\kappa'}})$.
\end{proof}

To certify that honest behavior is $\varepsilon$-optimal under a given
$\kappa$, we need an upper bound on $\OPT(M^{\mathcal{R}^s_\kappa})$.
We therefore use only the upper truncation $M^+$ when computing
$\kappa_{\min}$, and run at $N = 10^7$ samples per state.

\begin{figure}[tbp]
    \centering
    \begin{tikzpicture}
        \begin{axis}[
            paperplot, percent x,
            xlabel={$\alpha$},
            ylabel={$\kappa_{\min}$},
            xmin=0.045, xmax=0.105,
            ymin=0,
            legend pos=north west,
            xtick={0.05,0.06,0.07,0.08,0.09,0.10},
        ]
        \addplot[color=modelUniform,   mark=+, mark size=3.5pt, thick]  table[x=alpha, y=lin_uniform]   {figures/data/kappa.dat};
        \addplot[color=modelPerSlot,   mark=square]                     table[x=alpha, y=lin_perslot]   {figures/data/kappa.dat};
        \addplot[color=modelSmooth,    mark=triangle*] table[x=alpha, y=lin_smooth]    {figures/data/kappa.dat};
        \addplot[color=modelThreshold, mark=diamond*]  table[x=alpha, y=lin_threshold] {figures/data/kappa.dat};
        \legend{Uniform, Per-slot, Smooth, Threshold}
        \end{axis}
    \end{tikzpicture}
    \caption{Slashing threshold $\kappa_{\min}$ versus stake fraction
    $\alpha$, for each of the four MEV reward models of
    \cref{sec:results}.}
    \label{fig:kappa_min}
\end{figure}
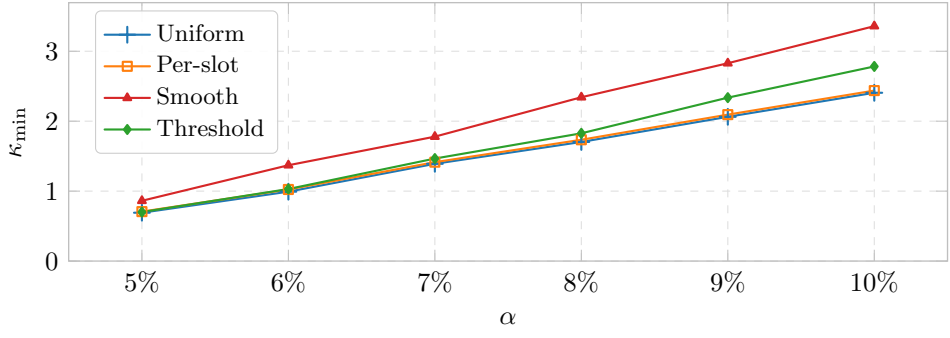
 
\subsection{Slashing Magnitudes}

\cref{fig:kappa_min} reports $\kappa_{\min}$ for the four MEV
reward models of \cref{sec:results} across stake fractions $\alpha$
from $0.05$ to $0.10$.
We observe that the ordering of $\kappa_{\min}$ at fixed $\alpha$
across MEV reward models matches the ordering of the manipulation
gain $\OPT(M^\mathcal{R}) - h^\mathcal{R}$ from \cref{sec:improvement}:
reward models that incentivize manipulation more strongly require
larger penalties to neutralize as expected. Similarly, $\kappa_{\min}$ is
monotonically increasing in $\alpha$, since larger adversaries gain
more from manipulation and require correspondingly larger slashing
magnitudes.

\subparagraph*{Solo-staker friendliness.}
One concern for introducing a missed slot penalty is the effect on
solo stakers.
Solo stakers in practice miss only a small fraction of their assigned
slots (roughly 1--2\%), as i.i.d.\ accidents rather than coordinated
drops at the tail. Under such a miss pattern, our slashing curve
is far weaker on such validators than a fixed per-slot penalty
that is independent of tail position and clustering. Misses early
in the epoch incur little or no penalty under the decaying weight
$S$, and the multi-miss multiplier stays close to $1$ for isolated
misses. A uniform per-slot penalty, by contrast, would charge the
same $\kappa$ for every accidental miss regardless of where in the
epoch it falls or how it clusters with other misses.
\section{Conclusion} \label{sec:conclusion}
We have shown that the incentive to manipulate RANDAO is highly dependent on
the MEV reward model. In particular, partial reward rollover and multi-block MEV both
amplify the manipulation gain, while lengthening the epoch is
a partial mitigation across all reward models we considered. We also showed that tail slashing, modeled as a reward function
within our methodology, can make honest behavior $\varepsilon$-optimal.
Our methodology can be used to analyze the incentives
around RANDAO in a variety of settings.
For instance, recent work \cite{cryptoeprint:2025/057} incorporates
a RANDAO bias analysis in the design of a protocol that uses randomness.
Our codebase and general approach can streamline
such analyses for other applications with complex reward models.

\subparagraph*{Future work.}
Three directions remain open.
First, \cref{sec:mitigation} analyzes a single tail-slashing curve
built from a decaying penalty and a multi-miss multiplier. A natural
extension is to explore alternative slashing curves, which may achieve
smaller costs for honest validators under the same deterrence guarantee.
Second, our framework works at the epoch level and we assume that the
adversary commits to an action for the entire epoch at once.
However, in practice, the adversary may do even better by adapting
their strategy based on the blocks they see in the current epoch online.
Analyzing this adaptive strategy is an interesting direction for
future work.
Third, the empirical gap between our upper and lower bounds is much
smaller than our sample-complexity bound predicts, suggesting room
for a tighter analysis.
 
    \section*{Acknowledgements}
    \paperacknowledgements

\begin{toappendix}
\section{Reward plots} \label{apx:reward-plots}
\begin{figure}[H]
    \centering
    \includegraphics[width=0.75\textwidth]{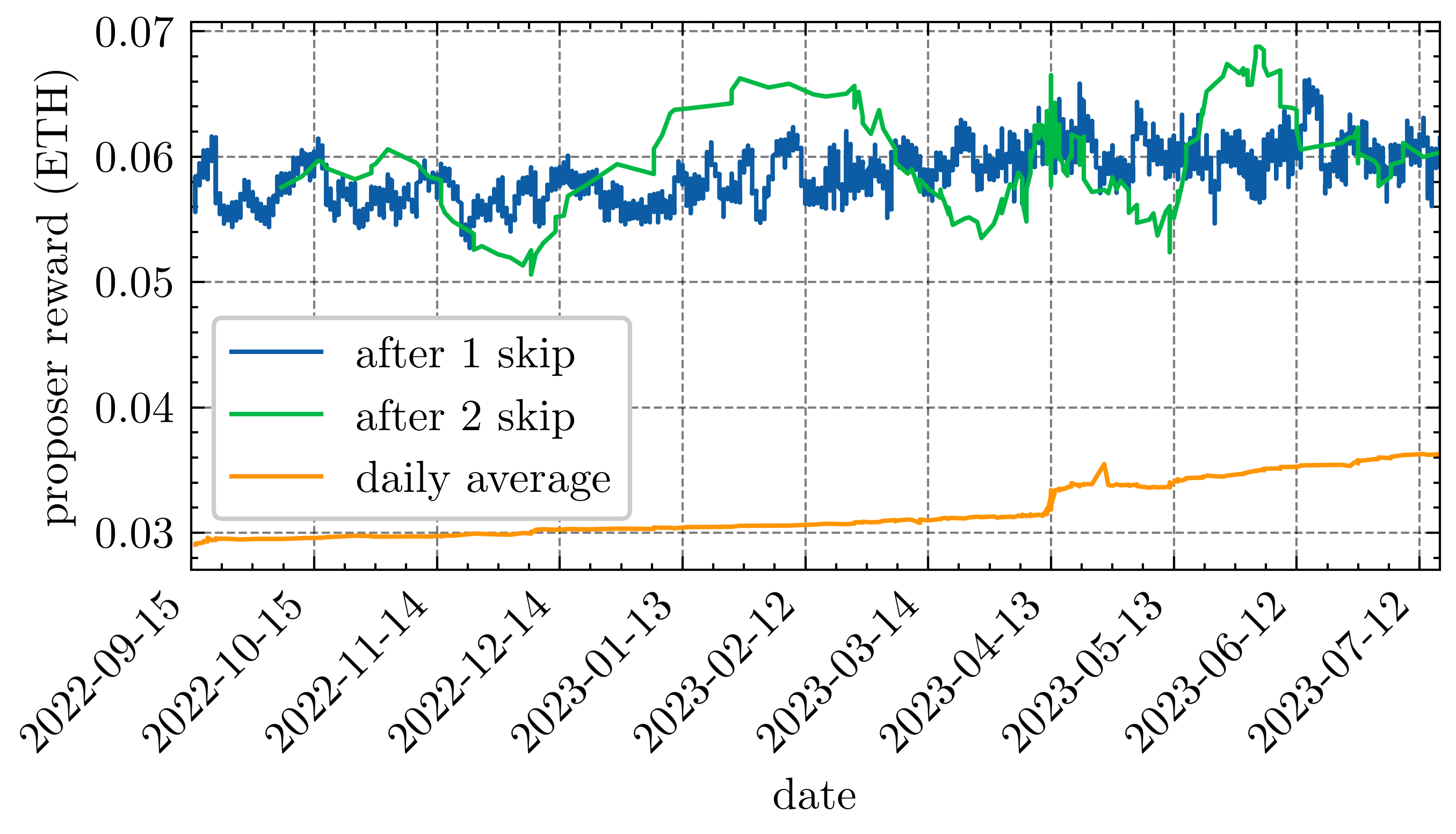}
    \caption{Consensus-layer rewards over time, excluding execution-layer
    rewards. Data from \cite{rewardsdataset}. After a single missed slot,
    the next block earns roughly twice the consensus reward. Two or more
    consecutive missed slots show no significant further increase.}
    \label{fig:cl_rewards}
\end{figure}

\begin{figure}[H]
    \centering
    \includegraphics[width=0.75\textwidth]{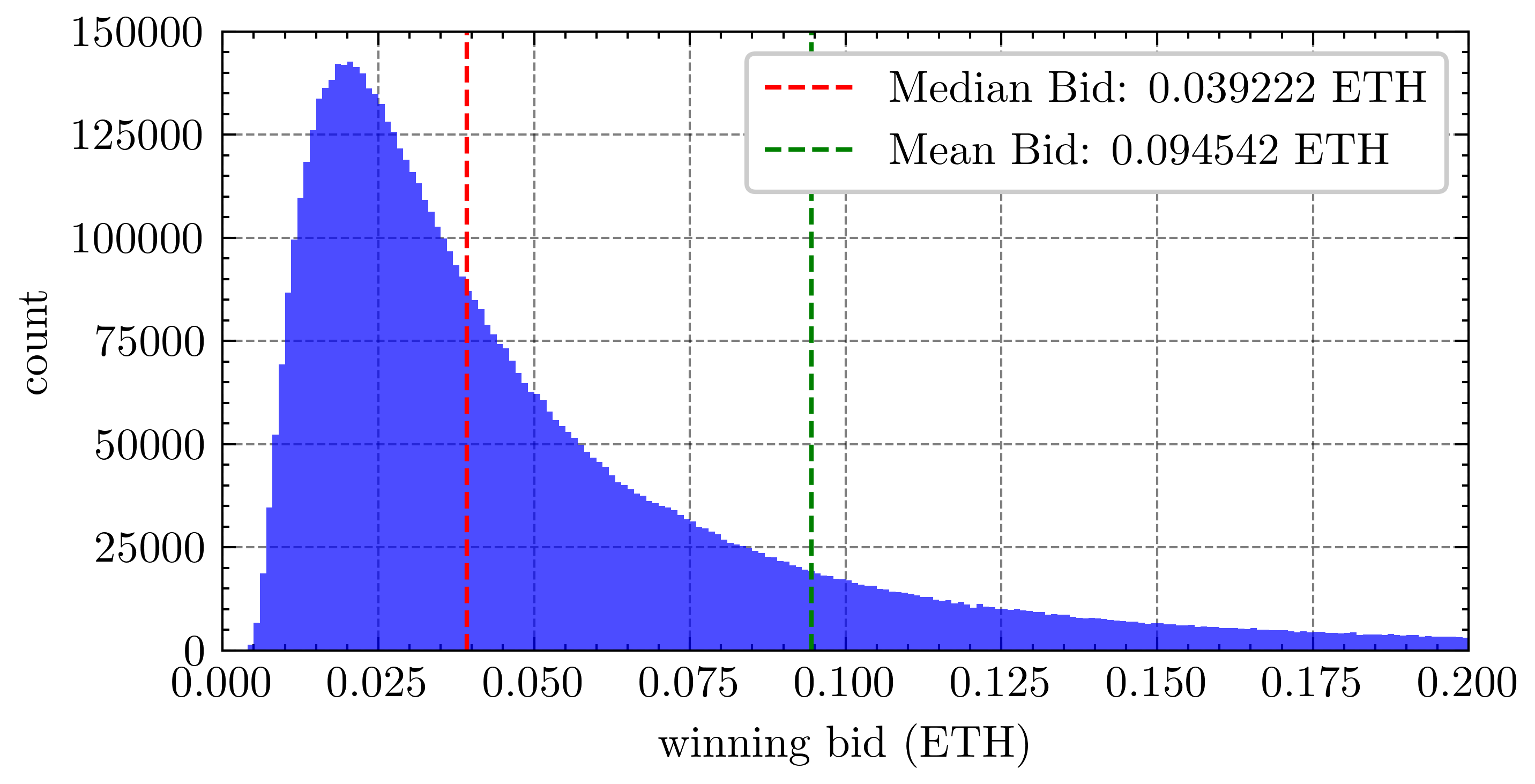}
    \caption{Histogram of bids from the MEV-Boost dataset \cite{dataalways}.}
    \label{fig:bid_histogram}
\end{figure}
\end{toappendix}

\bibliography{refs}

\end{document}